\documentclass[final,5p,times,twocolumn]{elsarticle}

\usepackage{amssymb, amsmath}
\usepackage{graphicx,cite,epsfig,amsfonts,multicol,subfigure,mathtools,bm,mathrsfs,setspace,color}
\usepackage{esint}
\usepackage{multirow}
\usepackage{times}
\usepackage{subfigure}
\usepackage{lipsum}
\usepackage{graphicx}
\usepackage{hyperref}
\hypersetup{hypertex=true,
	colorlinks=true,
	linkcolor=blue,
	anchorcolor=blue,
	citecolor=blue}

\journal{Physics Letters A}

\begin{document}
	
	\begin{frontmatter}
		
		
		
		\title{A Contradiction to the Law of Energy Conservation by Waves Interference in Symmetric/Asymmetric mode}

		\author{Bingli Jiao, Chenbo Wang, Zijian Zhou}
	    \address{School of Electronics, Peking University, Beijing, 100871, China}
		
		\begin{abstract}
			 It can be agreed that the linear superposition and energy conservation are two independent physics laws in general. The former allows the energy to be re-distributed over space and the latter restricts the energy in the total amount. However, Levine shows the contradiction of the two laws mentioned above by creating a cleaver model that demonstrates the energy ``doubling"- and ``missing" phenomenon with the constrictive- and destructive interference at every point of whole space, respectively. While, he presented a wrong explanation by using one of the radiating sources to compare with an isolated source by the compensation of the impedance, where the mistake is simply analyzed in this paper.  By setting up a spatial symmetric- and asymmetric-mode, we work upon Poynting theorem from the sources to the waves with the considerations of the superposition. The theoretical results reveal the invalidity of the energy conservation. Moreover, the experiments performed in the microwave anechoic chamber confirm the theoretical conclusion.
					 
		\end{abstract}

		
	\end{frontmatter}
	
	
	
	
	\section{Introduction}
	\label{introduction}
	It is well known that superposition of two waves having identical amplitudes and frequencies can produce a wave quadrupling the power density of a single wave or nullifying it by the in-phase- or $180^o$ out-phase interference, receptively, at some points in space. Let us refer the former to as the ``energy-doubling"- and the latter ``energy missing" behavior owning to the linear superposition of the waves after leaving their sources.

	In 1980, Levine created his model by placing two monochromatic radiation sources close to each other in the distance much smaller than the wave length, i.e., $d/\lambda <<1 $, for allowing the energy ``doubling" or ``missing" to occur approximately everywhere in whole space [1], where $d$ is the distance and $\lambda$ is the wave length. His explanation protects the energy conservation law by using the concept of wave-impedance without any detailed derivations. 
	
	Nonetheless, more researches were done either to make new explanations from different angles [2][3][4] or challenge the law of energy conservation continuously even shifted away from getting entangled with the sources [5].  
	
\begin{figure}[htbp] 
		\centering
		\includegraphics[width=0.8\linewidth]{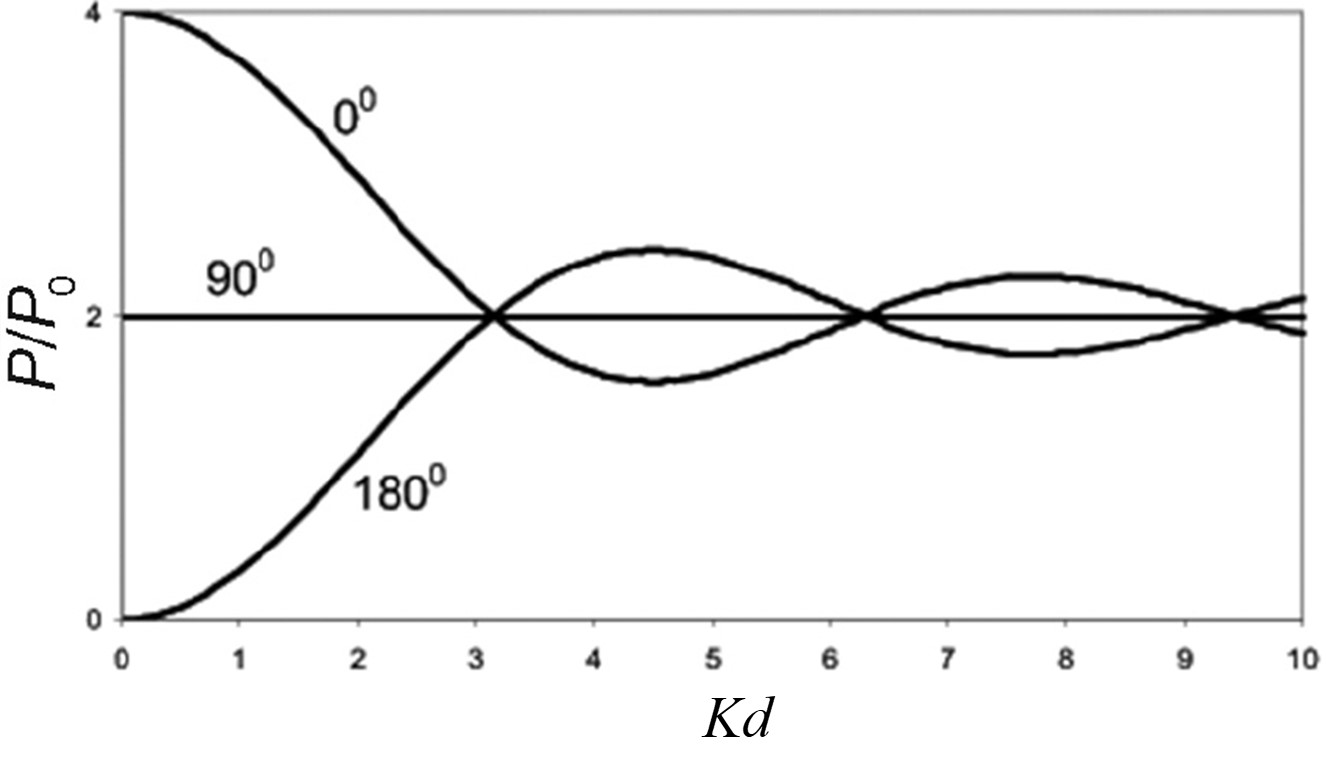}
		\caption{Normalized ($P/P_0$) power of radiation of two	coherent sound waves vs parameter for $0^o$, $90^o$, and $180^o$ phase shifts of oscillations of sound sources ($P$ is the power of radiation of two sound waves and $P_0$ is the audible power of an isolated source of sound, i.e., $P_0$ is the power generated by only one source in the absence of the second with the same amplitude of oscillations as for coupled oscillations).}
		\label{fig_lamda}
	\end{figure}

	One of the important studies is the work done by Drosd, Minkin and Shapovalov [6], who perform the simulations for investigating the total energy flows in Levine's model.  The results are ploted against the value of $Kd = 2\pi d/\lambda$ as shown in Fig. \ref{fig_lamda}, where ``power of radiation", i.e., the energy flow, with either the $0^o$ in-phase constructive or $180^o$ out-phase destructive interference varies with respect to $Kd$.  The results marked by $90^o$ obey the energy conservation law, because there is no interference between the two waves.  It is found that the energy conservation is the convergence-result of the interference when $d/\lambda$ goes to a large value.  
	
	Though the simulation results of Drosd and Shapovalov are obtained on the mechanical waves, we can take them as the extensions of Levine's phenomena (similar behaviors of the EM waves is available later in this paper).  In this regard, the mistake of the impedance-composition [1] can be immediately, because the reduction and increase of the total energy flow by the in-phase- and /$180^o$ out-phase interference, respectively, at $Kd = 5$ are precluded there.

\begin{figure}[htbp] 
		\centering
		\subfigure[]{
			\includegraphics[width=0.45\linewidth]{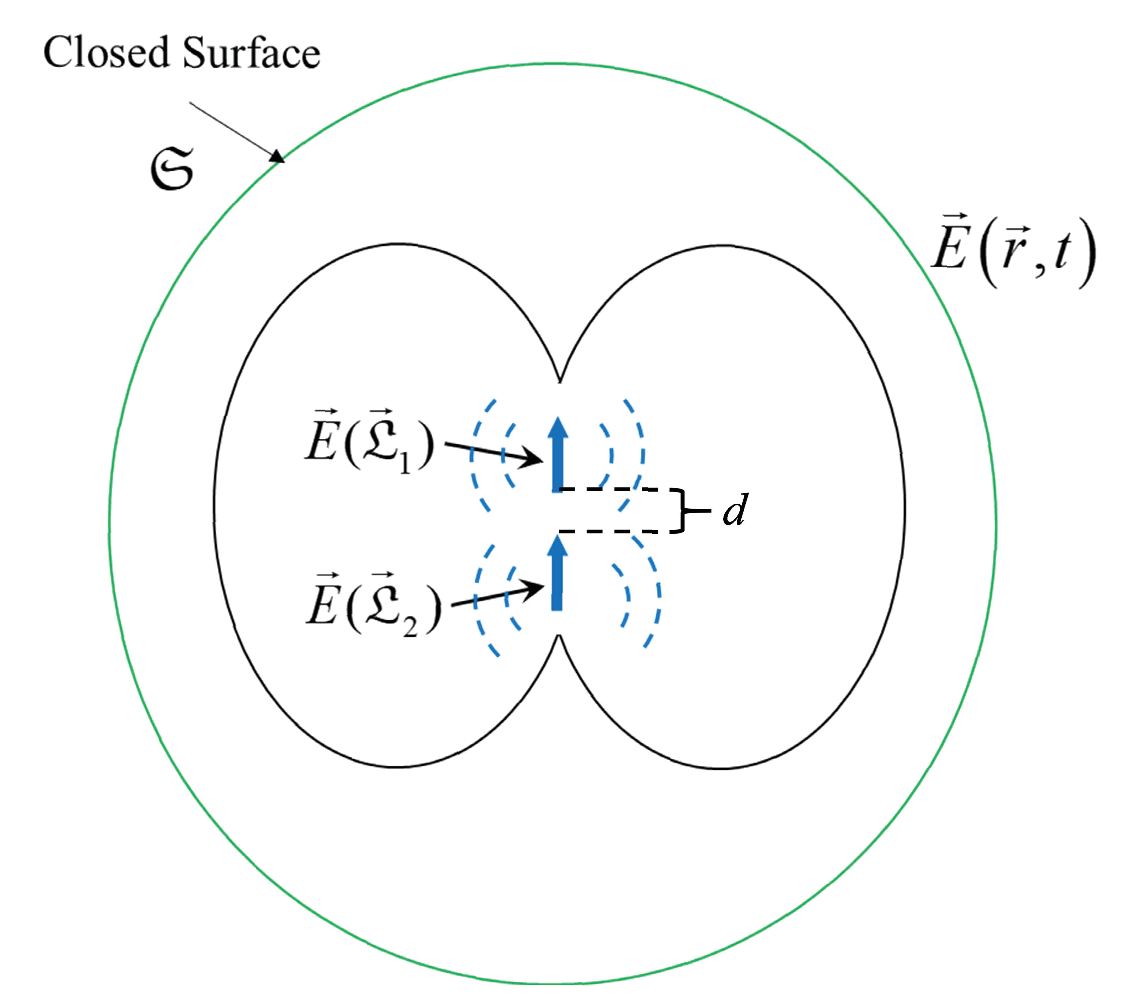}}
		\subfigure[]{
			\includegraphics[width=0.49\linewidth]{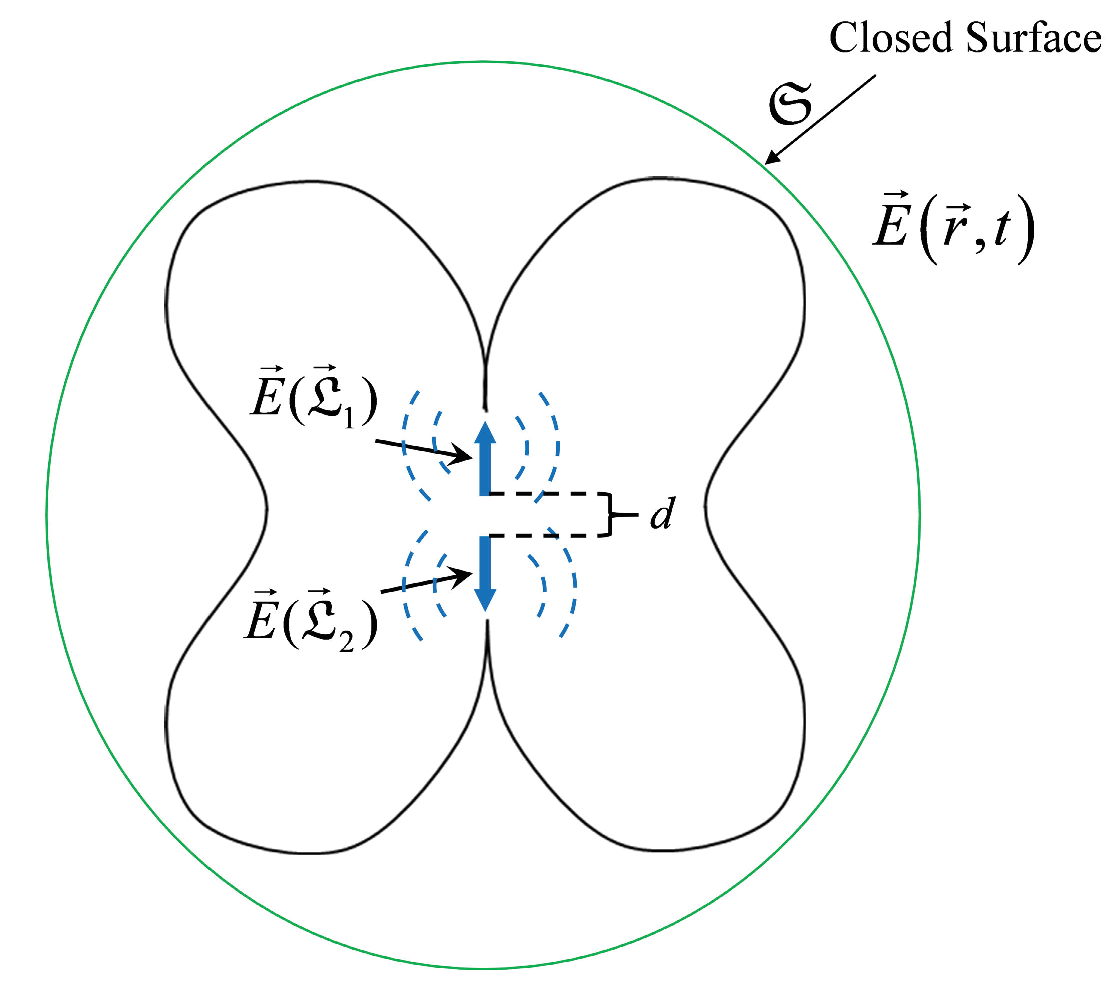}}
		\caption{The electric fields of the origin with S mode (a) and AS mode (b).}
		\label{fig_1}
	\end{figure}

	In fact, regardless of the coupling effects, we can apply the physics-thoughts of spatial symmetry/asymmetry to the two sources as shown in Fig. \ref{fig_1}, where (a) and (b) depict the symmetric mode (S-mode) and asymmetric mode (AS-mode), respectively.  Then, it is obviously found that the superposed amplitudes of the two EM waves of the S/AS mode can be approximately doubled/nullified everywhere over the entire space in Levine's model.  Based on the existence of the two modes,  the further theoretical derivations are carried out for revealing the physics-phenomena that are at odds with the energy conservation law in the scope of Poynting theorem.     
	
	To be conservative for drawing the surprising conclusion, the experiments are performed to confirm the theoretical approach.

	\section{Superposition Model}
	Let us consider a radiation system using two identical whip antennas to radiate monochromatic EM waves of the same frequency in free space. The length of each antenna is sufficiently small, compared to the wavelength, so that the exciting currents can be treated as the spatial-constant varying only with time, and dimension of the cross-section of each antenna is neglected in our research model.
	
	This work builds on the S- and AS mode as shown respectively in Fig. \ref{fig_1}(a) and (b) and starts directly on the total electric fields, i.e., the superposed electric fields   $\overrightarrow{E}(\overrightarrow{r},t)$ at the two antennas by  as
	\begin{equation}
		\begin{split}\label{01}
			\overrightarrow{E}(\overrightarrow{\mathfrak{L}}_1,t) = \pm  \overrightarrow{E}\overrightarrow{(\mathfrak{L}}_2,t) 
		\end{split}
	\end{equation}
	where $\overrightarrow{\mathfrak{L}}_1 = \{0,0,d/2\}$ and $\overrightarrow{\mathfrak{L}}_2 = \{0,0,-d/2\}$, and ``$\pm$" represents the S- and AS model, respectively.
	
	The densities of the currents at the both antennas can be expressed by 
	\begin{equation}
		\begin{split}\label{02}
			\overrightarrow{J}(\overrightarrow{\mathfrak{L}}_1,t) = \pm  \overrightarrow{J}\overrightarrow{(\mathfrak{L}}_2,t) 
		\end{split}
	\end{equation}     
	with $\overrightarrow{J} \equiv 0$ for $\overrightarrow{r} \notin (\overrightarrow{\mathfrak{L}}_1, \overrightarrow{\mathfrak{L}}_2)$, where $\overrightarrow{J}(\cdot)$ represents the density of the exciting current. 
	
	It is noted that though the lengths of the two antennas are neglected in expressions \eqref{01} and \eqref{02} for simplicity, the concept of current' s density remains when doing the integration at each antenna.     
	
	As long as the S/AS mode applies, the radiation power from each antenna is the same 
	\begin{equation}
	\begin{split}\label{05}
		\iiint\limits_{\mathfrak{L}_1}{{}} \left\langle \overrightarrow{J}\cdot \overrightarrow{E}\right\rangle dv 	= \iiint\limits_{\mathfrak{L}_2}{{}} \left\langle \overrightarrow{J}\cdot \overrightarrow{E}\right\rangle dv = -W/2
	\end{split}
\end{equation}	
and relates to the waves by Poynting's theorem    
	\begin{equation}
		\begin{split}\label{04}
			\mathop{{\int\!\!\!\!\!\int}\mkern-21mu\bigcirc}\limits_\mathfrak{S}
			\left\langle \overrightarrow{S}_i\right\rangle \cdot d\overrightarrow{s}+\iiint\limits_{\mathfrak{L}_i}{{}} \left\langle \overrightarrow{J}\cdot \overrightarrow{E}\right\rangle dv = 0 \ \ \ \ \ for \ \ \  \ \ \ i=1, 2
		\end{split}
	\end{equation}
	where $\overrightarrow{S}_i = \overrightarrow{E}_i\times \overrightarrow{H}_i$ represents the Poynting's vector, and $\overrightarrow{E}_i$ and $\overrightarrow{H}_i$ are the electric and magnetic fields owning to the radiation from antenna $i$ for $i=1,2$ (the electric fields of the S- and AS mode are shown in Fig. \ref{fig_1}),  $\mathfrak{S}$ denotes the closed surface that bounds each of the two sources, $\left\langle \cdot\right\rangle$ is the operator for time averaging, and $W/2$ equals the power provided by each source.
	
	\section{Total Radiation Power}
	Upon the supposition law, the total EM wave can be regarded as the linear superposition of the two EM waves, each of which owns to its source, i.e., owning to $\overrightarrow{E}(\overrightarrow{\mathfrak{L}}_i)$ and $\overrightarrow{J}(\overrightarrow{\mathfrak{L}}_i)$ for $i=1,2$, respectively. 
	
	The superposition law can be expressed as
	\begin{equation}
		\begin{split}\label{03-1}
			\overrightarrow{E} = \overrightarrow{E}_1 +\overrightarrow{E}_2 
		\end{split}
	\end{equation}   
	and
	\begin{equation}
		\begin{split}\label{03-2}
			\overrightarrow{H} = \overrightarrow{H}_1 +\overrightarrow{H}_2 
		\end{split}
	\end{equation}    
	where $\overrightarrow{E}_1$ and $\overrightarrow{H}_1$, $\overrightarrow{E}_2$ and $\overrightarrow{H}_2$ are the electric- and magnetic filed owing to radiations of antenna 1 and 2, respectively.

	By using the Poynting's integration with the consideration of \eqref{05}, we can calculate the total power carried by the superposed wave traveling to the outer space of $\mathfrak{S}$ 
	\begin{equation}
		\begin{split}\label{06}
			P_{AV} =\mathop{{\int\!\!\!\!\!\int}\mkern-21mu\bigcirc}\limits_\mathfrak{S}
			\left\langle \overrightarrow{S}\right\rangle \cdot d\overrightarrow{s} =  - W - \mathfrak{M}_{12}
		\end{split}
	\end{equation}
	with 
	\begin{equation}\label{07}
		\begin{split}
			\mathfrak{M}_{12} = \mathop{{\int\!\!\!\!\!\int}\mkern-21mu\bigcirc}\limits_\mathfrak{S}
			\left\langle {\overrightarrow{E}_1 \times \overrightarrow{H}_2 + \overrightarrow{E}_2 \times \overrightarrow{H}_1} 
			\right\rangle  \cdot d\overrightarrow{s} 
		\end{split}
	\end{equation}
	where $P_{AV}$ is the total power radiated to the space, $\overrightarrow{S} =\overrightarrow{E}\times \overrightarrow{H}$ represents the Poynting's vector of the superposed EM wave described in \eqref{03-1} \eqref{03-2} and $\mathfrak{M}_{12}$ is radiation power owning to the interference of the two waves.
	
	It is noted that the energy conservation requires $\mathfrak{M}_{12} \equiv 0$ in \eqref{06}.  However, a conflict is immediately found at the assumption of $l,d << \lambda$, because $\overrightarrow{E}_1 \approx \pm \overrightarrow{E}_2$ and $\overrightarrow{H}_1 \approx \pm \overrightarrow{H}_2$ hold in whole space, where ``$\pm$" represents the S- and AS mode, and $l$, $d$ and $\lambda$ represent the length of each antenna, distance between the two antennas and the wavelength, respectively.  
	
	The above approximations yield the results of \eqref{07} at 
	
	\begin{equation}\label{08}
		\begin{split}
			\mathfrak{M}_{12}  \approx \begin{cases} -W  \ \ \ \ \ \ \ \ \ \ \ \ \ \ \ \  \ \ \ \ \ \text{for}  \ \ \ S \ \text{mode} 
				\\ W \ \ \ \ \ \ \ \ \ \ \ \ \ \ \ \ \ \  \text{for}  \ \ \mathit{AS } \ \text{mode}
			\end{cases}
		\end{split}
	\end{equation}
	instead of zero.  The Levine's phenomena of ``energy doubling" and ``energy missing" present in form of 
	\begin{equation}\label{09}
		\begin{split}
			P_{AV} \approx \begin{cases} -2W  \ \ \ \ \ \ \ \ \ \ \  \ \ \ \  \ \ \text{for}  \ \ \ S \ \ \ \text{mode} 
				\\  \ \ \ \ 0  \ \  \ \ \ \ \ \ \ \ \ \ \ \ \ \ \ \ \ \  \text{for}  \ \ \ \mathit{AS } \ \text{mode}
			\end{cases}
		\end{split}
	\end{equation}
	which are the results owning to superposition of the two EM waves.  
	
	It can be agreed that law of energy conservation and that of the supposition are actually independent of each other, and the S- and AS mode show the comparability of the two laws at $d / \lambda << 1$. 
	
	The study on the problem of the general case for $d/ \lambda $ is presented in the next section. 		
	
	\section{Simulation and Discussion}
	This section presets the simulation results and an explanation to Levine' problem upon the concept of dipole's radiation.   
    \subsection{simulation result} 
	To confirm the theoretical results derived in the last section, simulations are performed in the system as shown in Fig. \ref{fig_1} by using the platform of Matlab (R2019a), where the monochromatic excitations of frequency 2.4 GHz are used to the two identical antennas, each length of which is $ 0.8$ cm. 
	
	For presenting the simulation results clearly, equation \eqref{06} is converted to 
	\begin{equation}\label{11}
		\begin{split}
			\beta= (P_{AV} + W)/W
		\end{split}
	\end{equation}	
	where $ \beta = \mathfrak{W}_{12}/W $ is defined as the interference factor. 
	
	\begin{figure}[htbp] 
		\centering
		\includegraphics[width=0.8\linewidth]{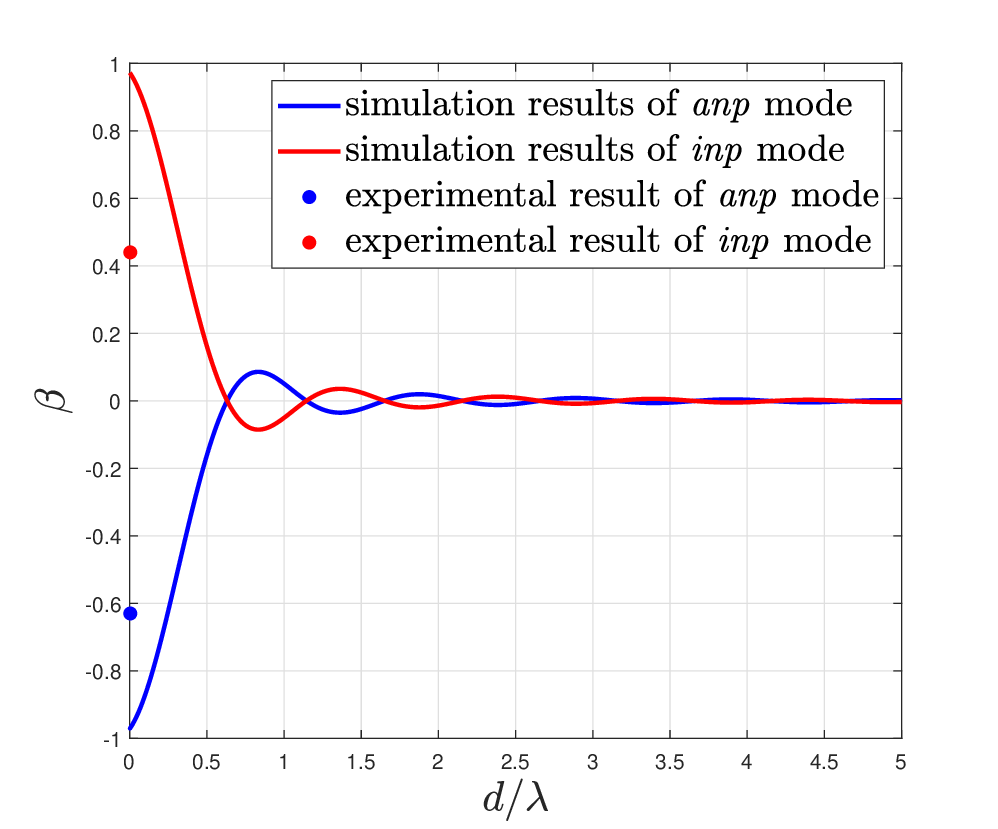}
		\caption{Simulated $\beta$ vs. $d/\lambda$ and the experimental results of S- and AS mode. }  
		\label{fig_2}
	\end{figure}
	
	It is noted that $\beta = 0$ in \eqref{11} is required for holding the energy conservation, while $\beta = +1$ and $\beta=-1$ indicate the energy ``doubling" and ``missing" respectively.  
	
	The simulations are performed for investigating the behaviors of $\beta$ against $d/\lambda$, where $d$ and $\lambda$ are the distance between the two antennas and the wavelength, respectively.  The results are shown in Fig. \ref{fig_2}, where one can infer that the energy conservation law holds for $ d/\lambda > 4 $ because of the rapid convergence with S- and AS mode.  
	
	It can be an important phenomenon that the total energy flow can fluctuate around the value required by the energy conservation, when $d/ \lambda $ is small.           
	
	In our simulations, the values of $\beta$ fluctuate around zero and finally approach $\beta=\pm 0.95$, respectively. The discrepancies of $\beta$ to the theoretical predictions at $ \pm 1$ are found due to the fact that the lengths of the two antennas are not infinitesimal as assumed in the theoretical model.   
	
	The simulated results above agree with those obtained by [6] in principle, which did not come to the theoretic derivations unfortunately.

	\subsection{Upon Dipole Model}
	This sub-section discusses the S- and AS mode at a limitation of $ d =  0$ by using the dipole model, i.e., $ql$cos$\omega t$, to each of the antennas, where $q$ is the electric charge, $l$ is the amplitude and  $\omega$ is the angel frequency and $t$ is the time.      
	
	Assuming that we are working in free space, the radiation power of one dipole has been calculated [7]   
	\begin{equation}\label{13-rad-0}
	P^0_{AV} = \frac{(P_0)^2 \mu_0\omega^4}{ 12\pi C}
	\end{equation}	
	with $P_0=ql$, where $P^0_{AV}$ is the radiation power of the dipole, C is the velocity of light and $\mu_0$ is the permeability of free space.   
	
	Now, we use the two dipoles to replace the two antennas and examine the radiation problem for $d \approx 0 $ as follows. 
	
		\begin{figure}[htbp] 
		\centering
		\includegraphics[width=1\linewidth]{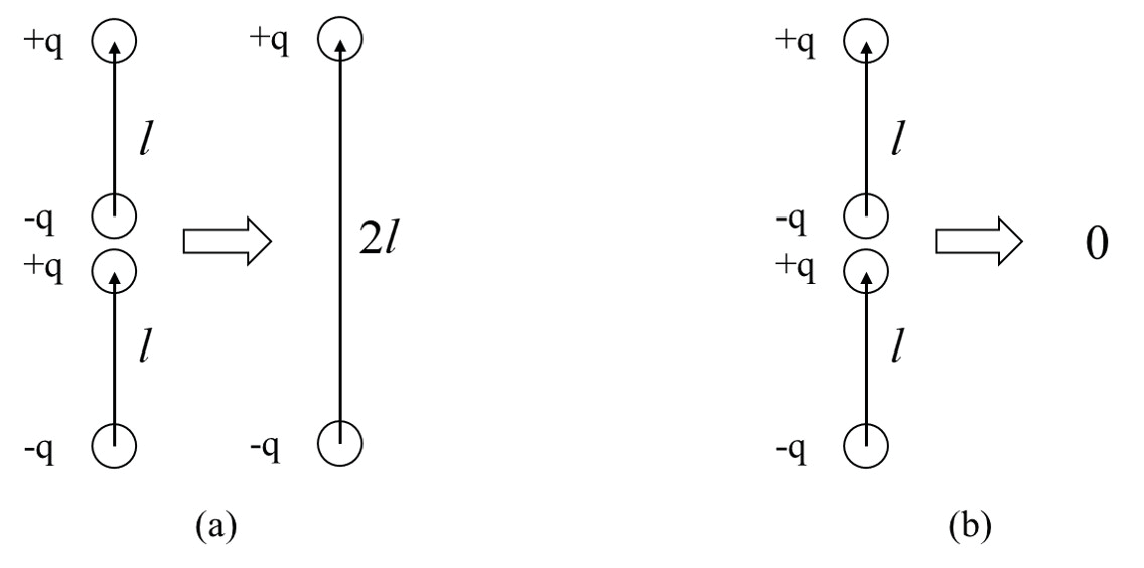}
		\caption{The two dipole-source modes: (a) S mode and (b) AS mode.}
		\label{fig_3-ab}
	\end{figure}
	
	For S mode, the two dipoles combine into one in form of  $P = q (2l)cos\omega t$ as shown in Fig. \ref{fig_3-ab}(a). The total radiation power of the combined dipole can be obtained by 
	\begin{equation}\label{13-rad}
	P^S_{AV} \approx \frac{(2P_0)^2 \mu_0\omega^4}{ 12\pi C} = 2\times (2P^0_{AV})
	\end{equation} 
    where $P^S_{AV}$ is the total power carried by the superposed EM wave, $(2P^0_{AV})$ is the power summation of the two dipoles and $2\times (2P^0_{AV})$ indicates the energy doubling phenomenon.
    
    	\begin{figure}[htbp] 
    	\centering
    	\includegraphics[width=0.8\linewidth]{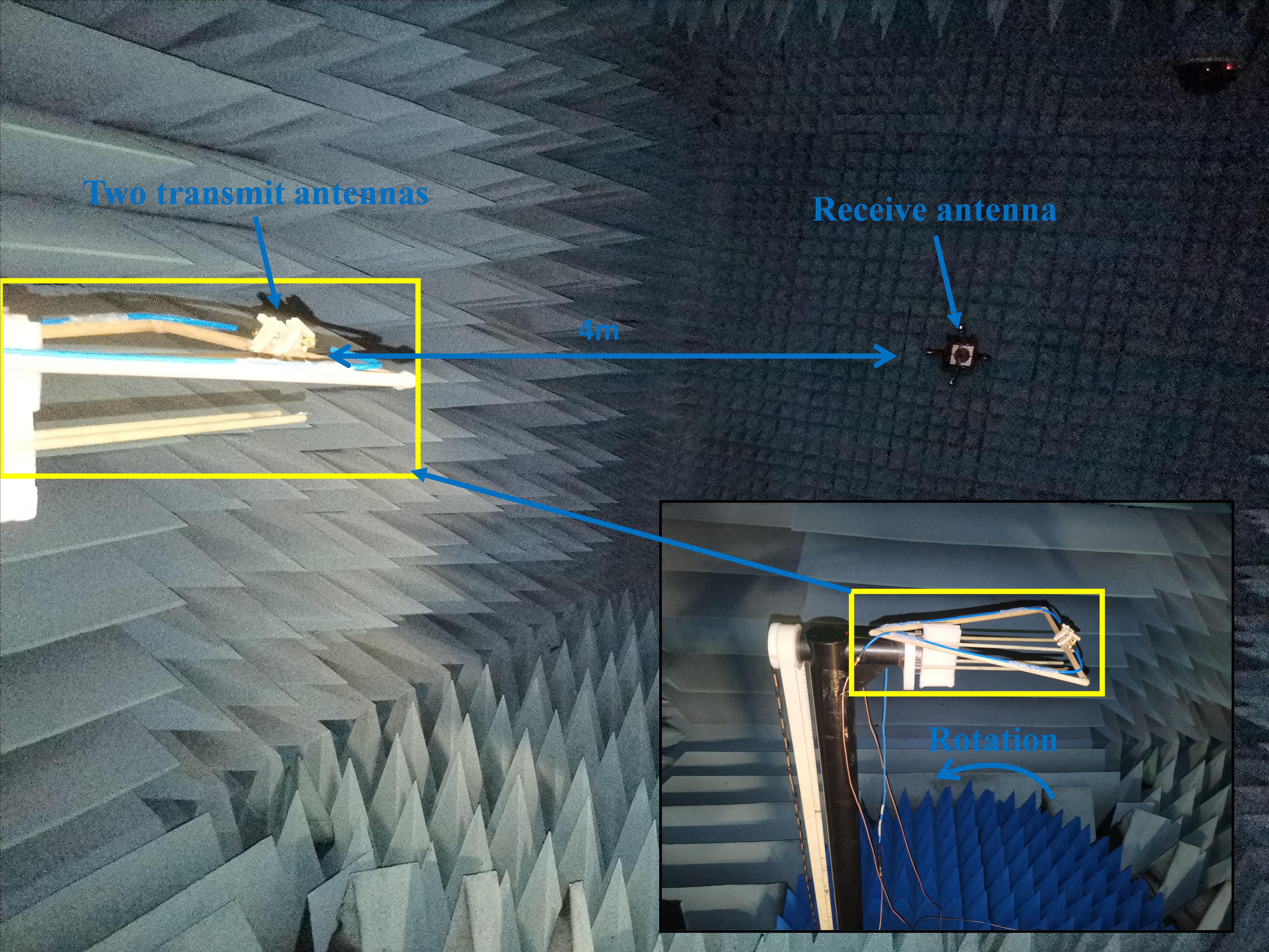}
    	\caption{The photo of the experimental environment and system.}
    	\label{fig_5S}
    \end{figure}
    
	While, for the AS mode, the two dipoles become a quadrupole as shown in Fig. \ref{fig_3-ab}(b), where the first order of the power-approximation can be expressed by  
\begin{equation}\label{13-rad-as}
P^{AS}_{AV} = O(P^0_{AV}).
\end{equation}
where $P^{AS}_{AV}$ is the total power carried by the superposed EM wave and $O(P^0_{AV})$ indicates the energy missing phenomenon in the AS mode. 
 
From the above derivations, it is found that though the two sources radiate the same power in the S/AS mode, the total power of the superposed EM wave does not equal to the power summation of the two sources, i.e., does not agree with the energy conservation.  

The authors note that the above conclusion holds in the scope of Poynting theorem.     
	
	\section{Experimental Confirmation}
	To confirm the theoretical- and simulation conclusions drawn in the system as shown in Fig. \ref{fig_1}, the experiments are performed as follows. 

	The total radiation power of the system are calculated by the experimental data of the power-patterns obtained in a microwave anechoic chamber (see Fig. \ref{fig_5S}) in sizes of 7.4m, 3.75m and 3.75m in length, width and height, respectively. The shielding effectiveness of vertical- and horizontal polarized EM wave are approximately 111.1dB and 111.5dB at radio frequency of 2.4GHz, respectively. The temperature and humidity in the chamber are controlled in the ranges from 22$C^o$ to 24 $C^o$ and from 41$\%$ to 45$\%$, respectively.
		
	\begin{figure}[htbp] 
		\centering
		\includegraphics[width=0.9\linewidth]{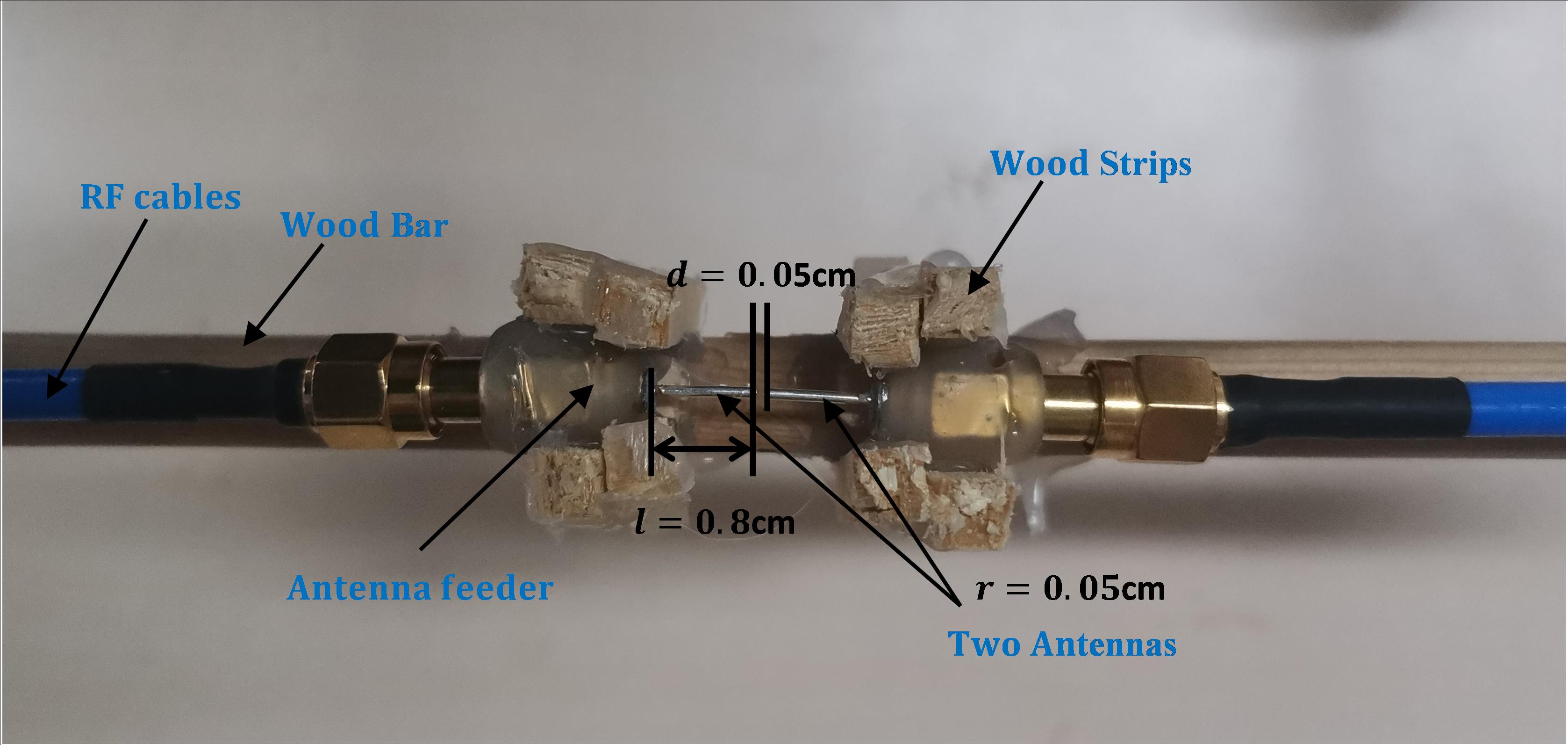}
		\caption{Photo of the structure of the two antennas.}
		\label{fig_3}
	\end{figure}		

	\begin{table*}[htbp] 
	\linespread{1.3}\selectfont
	\caption{The results of  $S$- and $AS$ mode from experimental results. }
	\centering
	\scalebox{0.8}[0.7]{
		\begin{tabular}{|ccccccccc|}
			\hline
			\multicolumn{9}{|c|}{The radiation powers from antenna 1 and 2 }                                                                                                                          \\ \hline
			\multicolumn{2}{|c|}{\multirow{2}{*}{}}                                          & \multicolumn{4}{c|}{Antenna 1}                                                                               & \multicolumn{3}{c|}{Antenna 2}                                                   \\ \cline{3-9} 
			\multicolumn{2}{|c|}{}                                                           & \multicolumn{1}{c|}{$W'_1=Z_{11}\left\langle I_1^2 \right\rangle$}       & \multicolumn{1}{c|}{$\eta_1 W'_1=Z_{21}\left\langle I_1I_2 \right\rangle$}  & \multicolumn{2}{c|}{$W_1$}    & \multicolumn{1}{c|}{$W'_2=Z_{22}\left\langle I_2^2 \right\rangle$} & \multicolumn{1}{c|}{$\eta_2 W'_2=Z_{12}\left\langle I_1I_2 \right\rangle$} & $W_2$    \\ \hline
			\multicolumn{2}{|c|}{$AS$ mode}                                        & \multicolumn{1}{c|}{0.0118mW}         & \multicolumn{1}{c|}{0.0013mW}        & \multicolumn{2}{c|}{0.0105mW} & \multicolumn{1}{c|}{0.0091mW}   & \multicolumn{1}{c|}{0.0010mW}       & 0.0081mW \\ \hline
			\multicolumn{2}{|c|}{$S$ mode}                                        & \multicolumn{1}{c|}{0.0092mW}         & \multicolumn{1}{c|}{0.0010mW}        & \multicolumn{2}{c|}{0.0102mW} & \multicolumn{1}{c|}{0.0073mW}   & \multicolumn{1}{c|}{0.0008mW}       & 0.0081mW \\ \hline
			\multicolumn{9}{|c|}{The experimental data for verifying \eqref{11}}                                                                                                                      \\ \hline
			\multicolumn{1}{|c|}{}                    
			&\multicolumn{3}{c|}{$W = W_1+W_2$}   
			&\multicolumn{3}{c|}{$P_{AVG}$}                                                                                   
			&\multicolumn{2}{c|}{$\beta$}                                                     \\ \hline				
			\multicolumn{1}{|c|}{$AS$ mode} 
			& \multicolumn{3}{c|}{0.0186mW}                                                
			& \multicolumn{3}{c|}{0.0069mW}                                        
			& \multicolumn{2}{c|}{-0.63}                                                       \\ \hline				
			\multicolumn{1}{|c|}{$S$ mode}  
			& \multicolumn{3}{c|}{0.0183mW}                                                
			& \multicolumn{3}{c|}{0.0263mW}                                        
			& \multicolumn{2}{c|}{0.44}                                                        \\ \hline		
	\end{tabular}}
\end{table*}

	\begin{figure}[htbp] 
	\centering
	\includegraphics[width=0.9\linewidth]{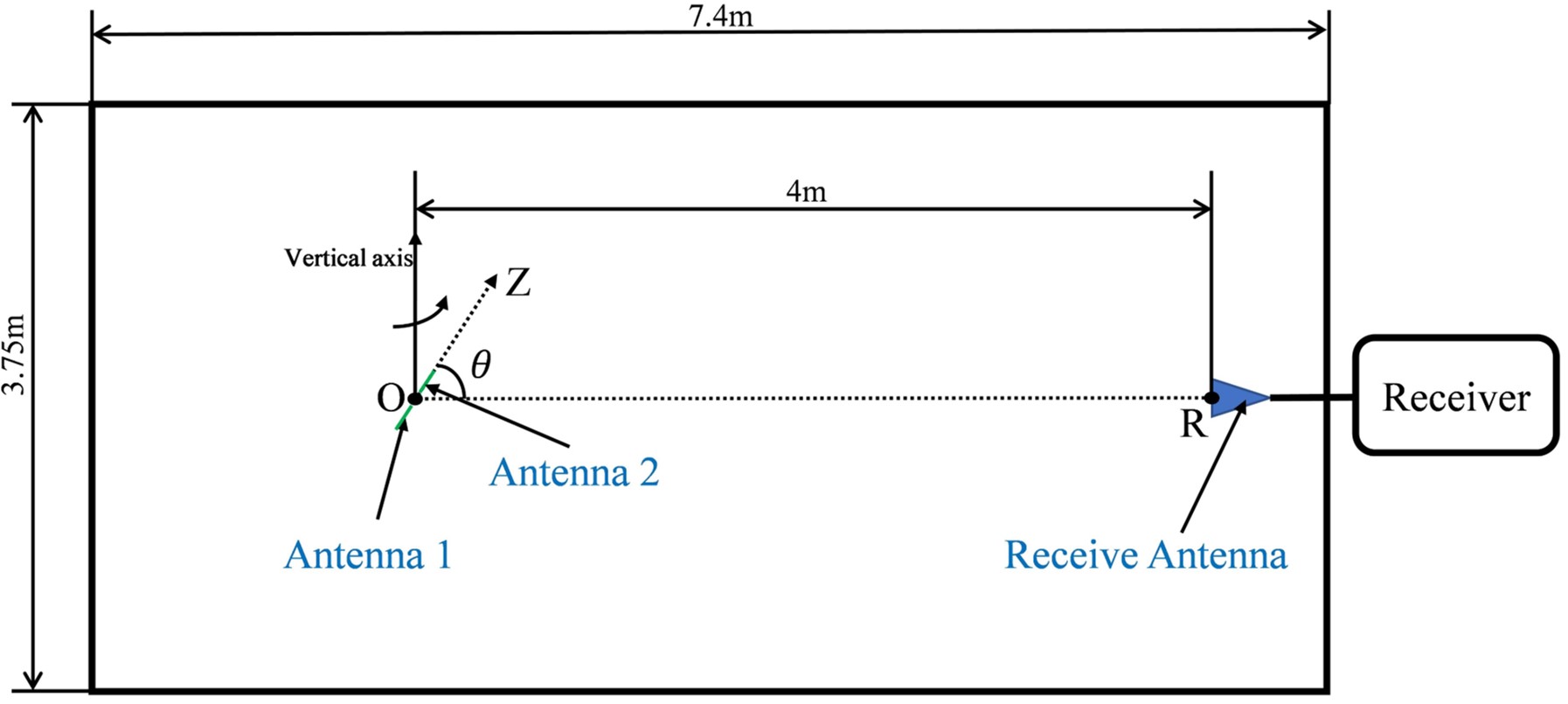}
	\caption{Rotation Planform.}
	\label{fig_4}
\end{figure}

	The two whip antennas are made using two cooper wires, of which the lengths and radii are the same at 0.8 cm and 0.05 cm, respectively.  By sticking the two antennas to a straight wooden bar, the geometric structure of the transmit antennas holds as shown in Fig. \ref{fig_3}.  For measuring the radiation power patterns, the line of the two antennas is arranged along with z-axis, which can rotate around its vertical axis as shown in Fig. \ref{fig_4},  where the angle between z-axis and line OR is marked by $\theta$ which can change, during the experiments, from $\theta=0^o$ to $360^o$ with angle resolution of one degree, i.e., stepped by $1^o$ at each measurement, and the distance from origin ``O" to ``R" is set at 4 meters for satisfying the far-field condition.  A  receive antenna is placed at position ``R" to record the radiation power.

	Once a power pattern has been measured over the range from $\theta=0^o$ to $360^o$, the Poynting integration can be calculated, because of the geometric symmetry around z-axis.

	Two Network Analyzers (Agilent ENA-L E5062A) are taken to the following uses: one works as the signal generator providing the signal-power to the two antennas and the other as the power-receiver.  	
	
\begin{figure}[htbp] 
		\centering
		\includegraphics[width=0.9\linewidth]{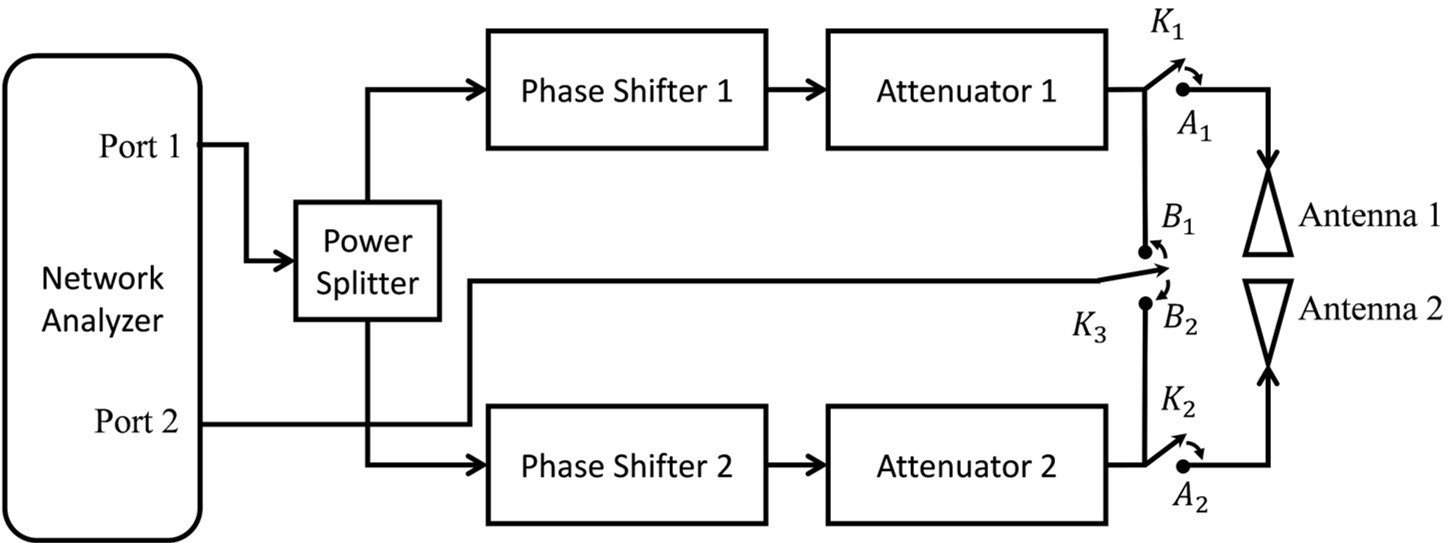}
		\caption{The circuit of adjusting the phase and power for realizing the $S-$ and $AS$ mode separately.}
		\label{fig_6S}
	\end{figure}

	Before doing the experiments for verifying the energy ``doubling"- and ``missing " effect,  we set up the $S$- and $AS$ mode by using the circuit as shown in Fig. \ref{fig_6S}.  The procedures of constructing the S mode are described as follows.  By opening switch $K_1$ and $K_2$ and closing $K_3$ to $B_1$, Port 1 of the Network Analyzer is used to input the sinusoidal signal to the circuit through the Power Splitter, Phase Shifter 1 and Attenuator 1 to Port 2 which records the power and phase of the signal as a reference signal.  Then, we keep $K_1$ and $K_2$ opened and close $K_3$ to $B_2$.  By adjusting Phase Shifter 2 and Attenuator 2 until the phase of received signal is $0^o$ with respect to and the power is same as those of the reference signal, the S mode has been constructed. Finally, by closing switches $K_1$ and $K_2$ and opening $K_3$ from $B_1$ and $B_2$, the radiations of the S mode can be realized.

	\begin{table*}[htbp] 
		\linespread{1.4}\selectfont
		\caption{EM coupling effect. }
		\centering
		\scalebox{1}[1]{
			\begin{tabular}{|ccccc|}
				\hline
				\multicolumn{1}{|c|}{\multirow{2}{*}{}} & \multicolumn{1}{|c|}{S matrix} & \multicolumn{3}{|c|}{ Z matrix and Coupling factor }                                                                                                                          \\ 
				\cline{2-5}
				\multicolumn{1}{|c|}{}&
				\multicolumn{1}{c|}{Voltage gains $S_{ij}$} &\multicolumn{2}{c|}{Calculated Impedances $Z_{ij}(\Omega)$}  &\multicolumn{1}{c|}{Coupling factor $\eta_{ij}$}                                                    
				\\ \hline
				\multicolumn{1}{|c|}{$(i,j)=(1,1)$}&
				\multicolumn{1}{|c|}{$0.630*e^{-122.0^{\circ}}$}&
				\multicolumn{2}{|c|}{$14.6-25.9j$}&
				\multicolumn{1}{|c|}{\multirow{2}{*}{$\eta_{12}=0.1068$}}
				\\
				\cline{1-4}
				\multicolumn{1}{|c|}{$(i,j)=(1,2)$}&
				\multicolumn{1}{|c|}{$0.040*e^{8.0^{\circ}}$}&
				\multicolumn{2}{|c|}{$1.56-1.15j$}&
				\multicolumn{1}{|c|}{\multirow{1}{*}{}}\\
				\hline
				\multicolumn{1}{|c|}{$(i,j)=(2,1)$}&
				\multicolumn{1}{|c|}{$0.040*e^{9.3^{\circ}}$}&
				\multicolumn{2}{|c|}{$1.59-1.12j$}&
				\multicolumn{1}{|c|}{\multirow{2}{*}{$\eta_{21}=0.1089$}}
				\\
				\cline{1-4}
				\multicolumn{1}{|c|}{$(i,j)=(2,2)$}&
				\multicolumn{1}{|c|}{$0.635*e^{-120.2^{\circ}}$}&
				\multicolumn{2}{|c|}{$14.6-26.9j$}&
				\multicolumn{1}{|c|}{\multirow{1}{*}{}}\\
				\hline
		\end{tabular}}
	\end{table*} 
	The AS mode can be realized in the similar manner above by adjusting the phase at $180^o$ with respect to the reference signal.

	Now, we calculate the results of \eqref{05} by using the experimental data of the first term on the left of \eqref{04}, where the coupling effects of the two sources are taken into account based on the microwave network theory [8] as follows.
	
	For the S- and AS mode, the radiation power of one antenna can be divided to two parts as

	\begin{equation}\label{13-05}
		W_i = Re\{Z_{ii} I^2_i \pm Z_{ij}I_iI_j\}\ \ \ \  \text{for} \ \ \ i,j = 1,2
	\end{equation}
	with 
		\begin{equation}\label{13-06}
			I_1= I_2
		\end{equation}
	where $W_i$ is the radiation power from antenna $i$ in presence of radiation of antenna $j$, $Z_{ii}$ and $Z_{ij}$ are the complex values of the radiation impedances of the transmission of antenna $i$ and that from antenna $i$ to $j$, $\left\langle \cdot \right\rangle$ and $Re\{\cdot\}$ are the operators of time averaging and taking the real value, ``$\pm$" represents the $S$- and $AS$ mode and $I_i$ and $I_j$ are currents in antenna $i$ and $j$ , respectively.  
	
	Because of \eqref{13-06}, \eqref{13-05} can be written as 
	\begin{equation}\label{13-07}
		W_i = W'_iRe\{1 \pm  \eta_{ij})\} \ \ \ \  \text{for} \ \ \ i,j = 1,2
	\end{equation}
	with  $  W'_i = Z_{ii} I_i^2  $, where $W'_i$ is the radiation power of antenna $i$ without consideration of the coupling effect and $\eta_{ij} = Re(Z_{ij})/Re(Z_{ii})$ is defined as the coupling factor. 
	
    To begin with experimental work on \eqref{13-07},  we use the Poynting integration to obtain $W'_i$ by 
    \begin{equation}\label{15}
    	W'_i = \varoiint\limits_\mathfrak{S}\left\langle \overrightarrow{S}'_i\right\rangle \cdot d\overrightarrow{s} 
    \end{equation}
    where $\left\langle \overrightarrow{S}'_i\right\rangle \cdot d\overrightarrow{s}$ is the experimental data of the received power measured from $\theta=0^o$ to $360^o$ by using antenna $i$ to transmit power without participation of the other antenna.  The values of $W'_1$ and $W'_2$ of the S- and AS model are listed in Table 1. 
	
	While, to complete the measurement of $W_i$, we need calculate impedances $Z_{ii}$ and $Z_{ij}$ by using the experimental data of $S_{ij}$ 
	\begin{equation}
		\left[ \begin{matrix}
			Z_{11} & Z_{12}  \\
			Z_{21} & Z_{22} \\
		\end{matrix} \right]={{Z}_{0}}\left( \left[ \begin{matrix}
			1 & 0  \\
			0 & 1  \\
		\end{matrix} \right]+\left[ 
		\begin{matrix}
			{{S}_{11}} & {{S}_{12}}  \\
			{{S}_{21}} & {{S}_{22}}  \\
		\end{matrix} \right] \right){{\left( \left[ \begin{matrix}
					1 & 0  \\
					0 & 1  \\
				\end{matrix} \right]-\left[ \begin{matrix}
					{{S}_{11}} & {{S}_{12}}  \\
					{{S}_{21}} & {{S}_{22}}  \\
				\end{matrix} \right] \right)}^{-1}}
	\end{equation}
	where  $Z_0=50 \Omega$ is the characteristic impedance, $S_{11}$ is the input port voltage reflection coefficient,  $S_{12}$ is the reverse voltage gain, $S_{21}$ is the forward voltage gain and $S_{22}$ is the output port voltage reflection coefficient defined in [7].
	
	\begin{figure}[htbp] 
		\centering 
		\includegraphics[width=0.6\linewidth]{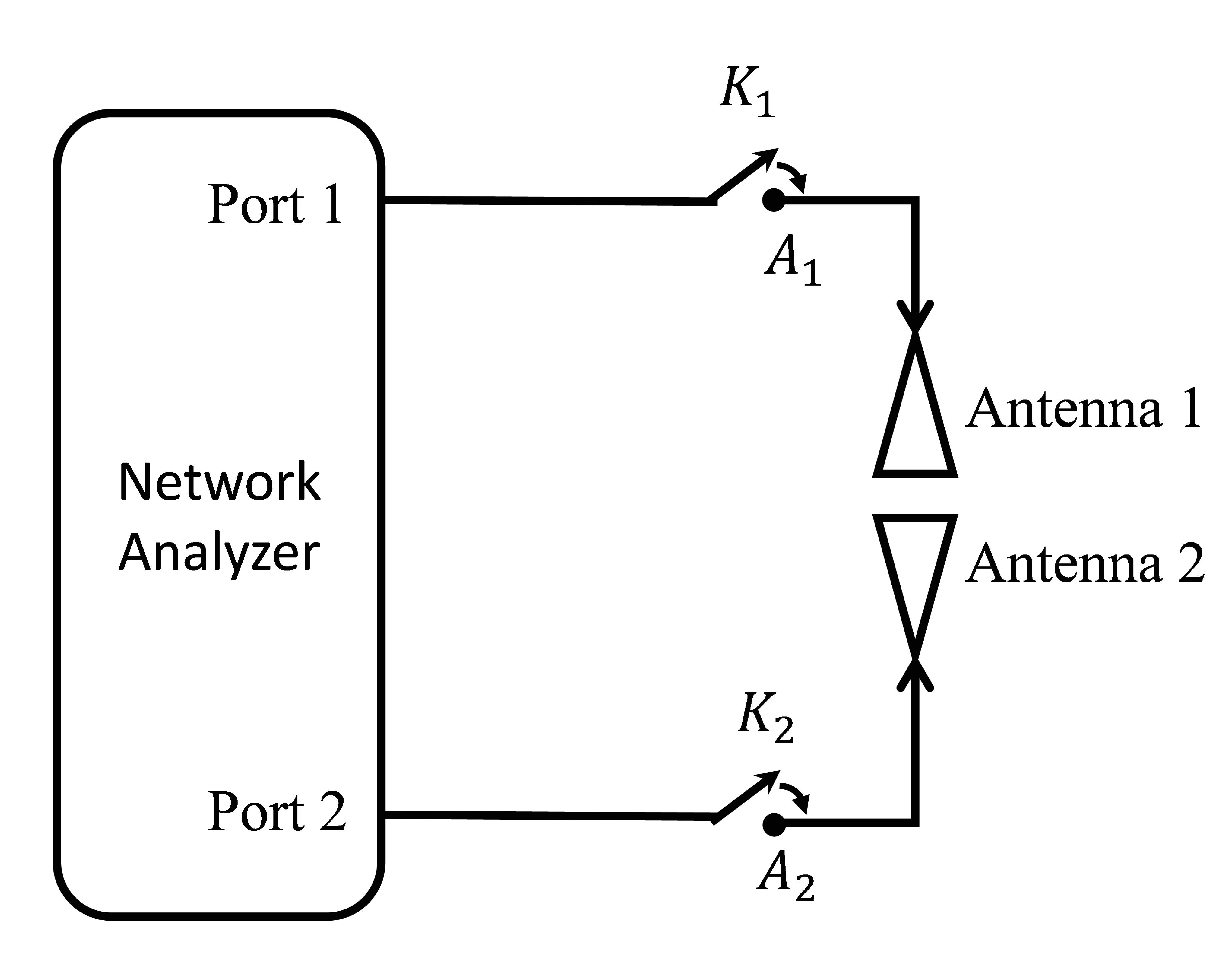}
		\caption{Measurement of each antenna response.}
		\label{fig_7S}
	\end{figure}

	Since the Phase Shifter 1 and 2 (see Fig.\ref{fig_6S}) place the roles as power-valves, we use the circuit as shown in Fig. \ref{fig_7S} to measure $S_{ij}$ as follows.  For measuring  $S_{ii}$, Port $i$ is used to record the value by closing $K_i$ and opening $K_j$.  Then, by closing $K_1$ and $K_2$, we use Port $j$ as the signal transmitter and Port $i$ as the receiver, $S_{ij}$ is recorded for $i,j = 1 ,2$ and $i \ne j$, respectively.   
	
    The values of $S_{ij}$ are listed in Table 2, where the results of the calculated $Z_{ij}$ and $\eta_{ij} =  Re\{Z_{ij}\}/Re\{Z_{ii}\}$ of the S- and AS mode are also listed.
	
	It is noted that S- and AS mode are approximated by the experimental values of $W_1$ and $W_2$ as shown in Table 1, which slightly deviated from the ideal values at about $20\%$.  Since the two values are measured by the radiation power, the $20\%$ deviation do not over turn the experimental evidence at $40\%$ unbalanced power for holding energy conservation (see the final results latter in last two rows in Table 1).  
	
	\begin{figure}[htbp] 
		\centering
		\includegraphics[width=0.8\linewidth]{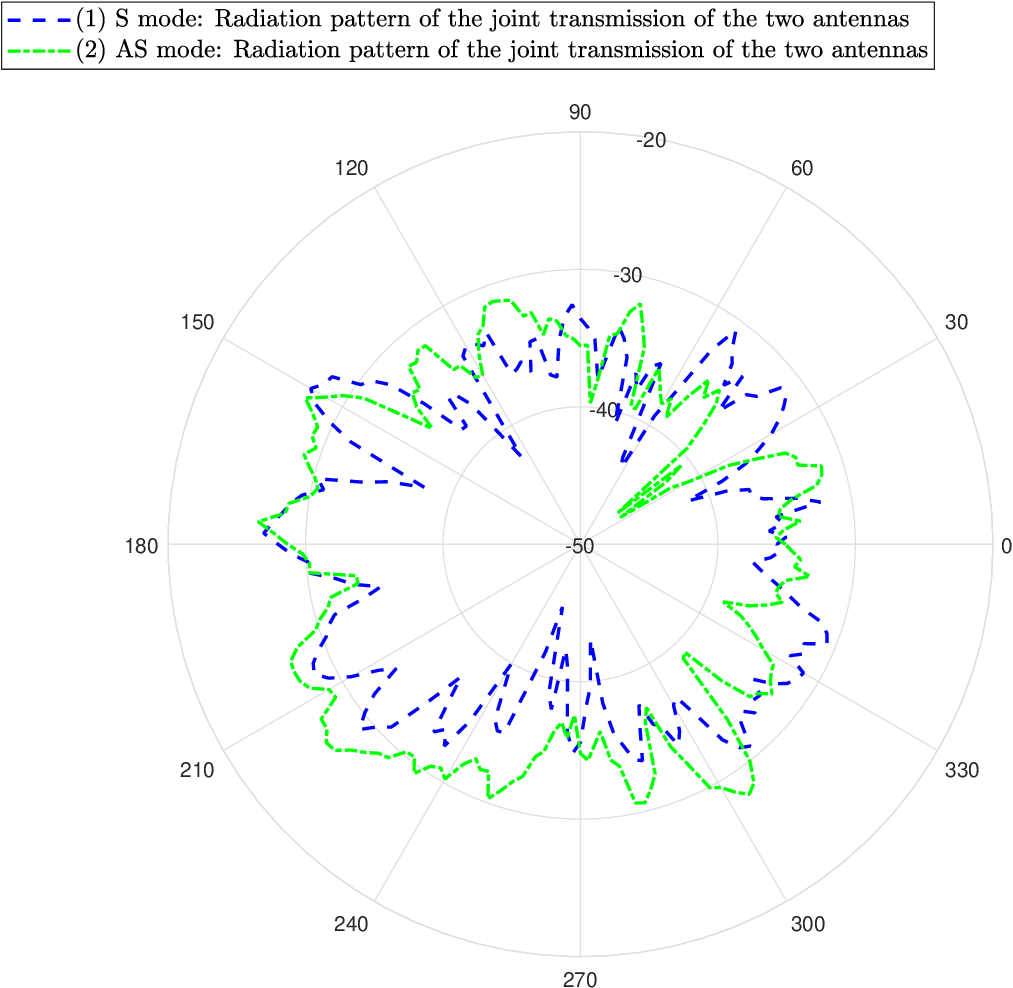}
		\caption{ 
			(1) Radiation patterns of the joint transmission of the two antennas in the $S$ mode measured in decibel,
			(2) Radiation patterns of the joint transmission of the two antennas in the $AS$ mode measured in decibel.}	
		\label{fig_8}
	\end{figure} 
	
	Finally, the radiation powers of the superposed EM waves were recorded by closing $K_1$ and $K_2$ with $K_3$ opened for the S- and AS mode separately.  The radiation patterns are shown in Fig. \ref{fig_8} for the two modes. The results of integral $P_{AVG} = \varoiint\limits_\mathfrak{S}\left\langle \overrightarrow{S}\right\rangle \cdot d\overrightarrow{s}$ are listed in Table 1, where one can find that $P_{AVG} > W_1 + W_2$ in the S mode and $P_{AVG} < W_1 + W_2$ in the AS mode, respectively.  In addition, the values of interference factor $\beta$ are also obtained by the experiments as shown in Table 1 and plotted in Fig. \ref{fig_2} for the comparisons with the simulation results.
	
	All results of this work conform the energy doubling- and missing phenomena.

	\section*{Acknowledgments}
	The experiments are taken in the microwave anechoic chamber of TsingHua University. 
	
	
	
	

	
	
	

	
\end{document}